# A robust near-field body area network based on coaxially-shielded textile metamaterial

Xia Zhu[1,2], Ke Wu[1,2], Xiaohang Xie[1,2], Stephan W. Anderson[2,3], and Xin Zhang[1,2*]

[1]Department of Mechanical Engineering, Boston University, Boston, MA 02215, United States.

[2]Photonics Center, Boston University, Boston, MA 02215, United States.

[3]Chobanian & Avedisian School of Medicine, Boston University Medical Campus, Boston, MA, 02118, United States.

*Corresponding author. E-mail: xinz@bu.edu

Contributing authors: xiaz@bu.edu; wk0305ok@bu.edu; xhxie@bu.edu; sande@bu.edu

**Abstract**

A body area network (BAN) involving wearable sensors populated around the human body can continuously monitor physiological signals, finding applications in personal healthcare and athletic evaluation. Existing near-field communication (NFC)-enabled BAN solutions, while facilitating reliable and secure interconnection among battery-free sensors, face challenges such as limited spectral stability against external interference. Here we demonstrate a textile metamaterial featuring a coaxially-shielded internal structure designed to mitigate interference from extraneous loadings. The metamaterial can be patterned onto clothing to form a scalable, customizable network, enabling communication between NFC-enabled devices and developed battery-free textile NFC sensing nodes placed within the network. Proof of concept demonstration shows the metamaterial's robustness against mechanical deformation and exposure to lossy, conductive saline solutions, underscoring its potential applications in wet environments, particularly in athletic activities involving water or significant perspiration, offering insights for the future development of radio frequency components for a robust BAN at the system level.

**Key words**

**Introduction**

The recent advancements in wearable technologies, encompassing physiological sensors[1-6], wireless communication protocols[7-9], and low-power electronics[10-12] have made the widespread adoption of scalable body area network (BAN) an increasingly realistic scenario. A typical BAN refers to a wearable wireless network interconnecting discrete physiological sensors distributed across the human body[13,14]. These sensors continuously capture and stream physiological changes or environmental variations to a central data hub, such as a smartphone, positioned in clothing pockets or surface-mounted onto the human body. Simultaneous monitoring of physiological activities from different anatomical regions holds great significance and finds applications in healthcare systems, including rehabilitation, chronic disease detection, gait analysis and athletic performance evaluation[15-17].

The stable, secure and uninterrupted transfer of sensor data within the BAN relies primarily on the interconnectivity between multiple sensors and the central data hub. While sensors can be directly hardwired to the data hub in clinical scenarios, wireless communication technique are essential for BANs to be compatible with and unhindered by users' daily activities. Popular wireless communication technologies employed in BANs include Bluetooth, Zigbee and Wi-Fi[18,19]. These radio frequency (RF) techniques facilitate a long-range far-field wireless connection with individual sensors, but they typically require power from an external battery or additional energy harvester for each sensing node, limiting user comfort while adding complexity in monitoring battery status and replacement[20,21]. Furthermore, far-field RF-enabled BANs exhibit inefficiency in signal transmission due to high radiation loss through the transmission media, demanding higher power input. Moreover, these approaches offer limited data security and require additional encryption effort[22]; otherwise, the user's privacy may be compromised as an unauthorized third-party polling device may easily intercept user data when placed adjacent to the BAN.

To address the previously mentioned challenges, recent studies propose BANs based on near-field communication (NFC)[23-26], an RF identification technique widely supported by most smartphones. Unlike radiative transmission, NFC operates in the near-field region adjacent to the transmitting/receiving antenna, where the electromagnetic field is predominantly non-propagating. NFC allows for the powering of multiple battery-free NFC tags through a reader while continuously retrieving data from these tags via inductive coupling. NFC typically has a short communication range of only a few centimeters, and while this characteristic results in robust and secure data transmission that can prevent eavesdropping, an additional near-field-enabled pathway is required to extend the NFC communication range of the BAN to the scale of the human body.

Integrating NFC technology into electronic textiles provides a seamless and non-invasive connectivity solution within the BAN, offering communication pathways that connect several distant hotspots (ideal sensor locations) to a central data hub[24,26]. The technique involves computer-aided digital embroidery of commercial conductive thread or liquid metal fiber. However, this method includes an irreversible sewing process with predetermined sensor terminal locations, rendering it unsuitable for user-customizable designs with specific target locations for the sensor terminals. Additionally, this approach exposes the embroidered RF component directly to the human body and the environment. Since the separation between the BAN and the human body is highly dynamic and variable, and users may frequently encounter high-permittivity and conductive media during daily activities, there exists a risk of spectral instability and predominant dielectric loss—a common issue encountered in various RF BAN scenarios[27-29]. These concerns were addressed in a textile-integrated resonating loop array (metamaterial), where the NFC signal may be transmitted through inductive coupling along the resonator array in the form of a magneto-inductive wave, enabling connectivity with multiple sensors placed at arbitrary positions along the BAN[25]. This approach involves a resonator design cut from metal foil and a slotted ground metal foil layer, integrated onto pre-existing clothing. While the design demonstrates moderate spectral stability on just one side, it remains sensitive to more extreme loading, such as water. In addition, the mechanical durability of the slotted metal foil design remains questioned during daily wear.

In this paper, we introduce a highly customizable BAN composed of textile metamaterial patch units, capable of operating via near-field communication to continuously power and retrieve physiological signal from battery-free NFC sensing nodes populated along the BAN (Fig. 1a). We draw inspiration from recent developments in biomedical imaging[30,31], adopting a meticulously designed resonator structure, crafting the metamaterial directly from off-the-shelf micro coaxial cable. By employing computer-aided digital embroidery technology, the metamaterial units are seamlessly integrated into a substrate textile, allowing attachment to pre-existing clothing in a user-defined customized pattern. The network facilitates communication between an NFC reader/NFC-enabled smartphone and multiple sensors placed in proximity to the network. In sharp contrast to previous efforts to develop NFC-enabled wearable BANs[24-26], the metamaterial proposed herein demonstrate remarkable spectral stability and insensitivity against extraneous loadings. We demonstrate the network's robustness against environmental changes induced by human motion and its maintained functionality when exposed to extreme conditions, such as being immersed in saline solutions with concentrations up to seawater levels. This resilience indicates its durability against washing cycles, rainy environments and perspiration, suggesting potential applications in body-tracking systems and athletic performance evaluation in water sports. This unique characteristic is primarily attributed to the metamaterial's internal coaxial structure, inherited from the constituent micro coaxial cable, effectively constraining the electric field within, while mitigating capacitive coupling with the surrounding environment. Additionally, we develop textile-integrated NFC sensing nodes compatible with commercially available NFC readers and NFC-enabled smartphones. These sensing nodes can interface with a variety of analog sensors through a commercial NFC transponder, allowing placement or attachment to various anatomical locations on clothing based on user demands, enabling a scalable BAN that facilitates wireless, continuous physiological signal monitoring at a system level.

## Results

### Construction of the metamaterial

Magnetic metamaterials typically comprise sub-wavelength resonating units with various configurations, such as capacitively loaded single-turn loops or spiral/helical inductor coils[32-34]. In our design, we employ a two-turn closed-loop resonator crafted from commercially available micro coaxial cable with a diameter of 0.99mm (Fig. 1b, Supplementary Fig. 1a). Employing a computer-aided digital embroidery pattern, two layers of textile are securely bonded together, creating a textile duct where the metamaterial is then inserted and assembled. The coaxial cable adopted herein consists of four layers: the inner conductor (silver-plated copper braid), perfluoroalkoxy (PFA) insulation, outer conductor (silver plated copper braid shield) and PFA shield. This inherently thin and flexible coaxial cable imposes no constraints on human activities when integrated into the textile BAN. The metamaterial unit cell is crafted by welding the inner conductor at one end of a cable segment to the outer conductor at the opposing end at the open slit. One end of the inner conductor is left hanging and open-circuited to generate resonating current upon an incoming excitation. The open slit is then reinforced with a waterproof heat shrink tubing.

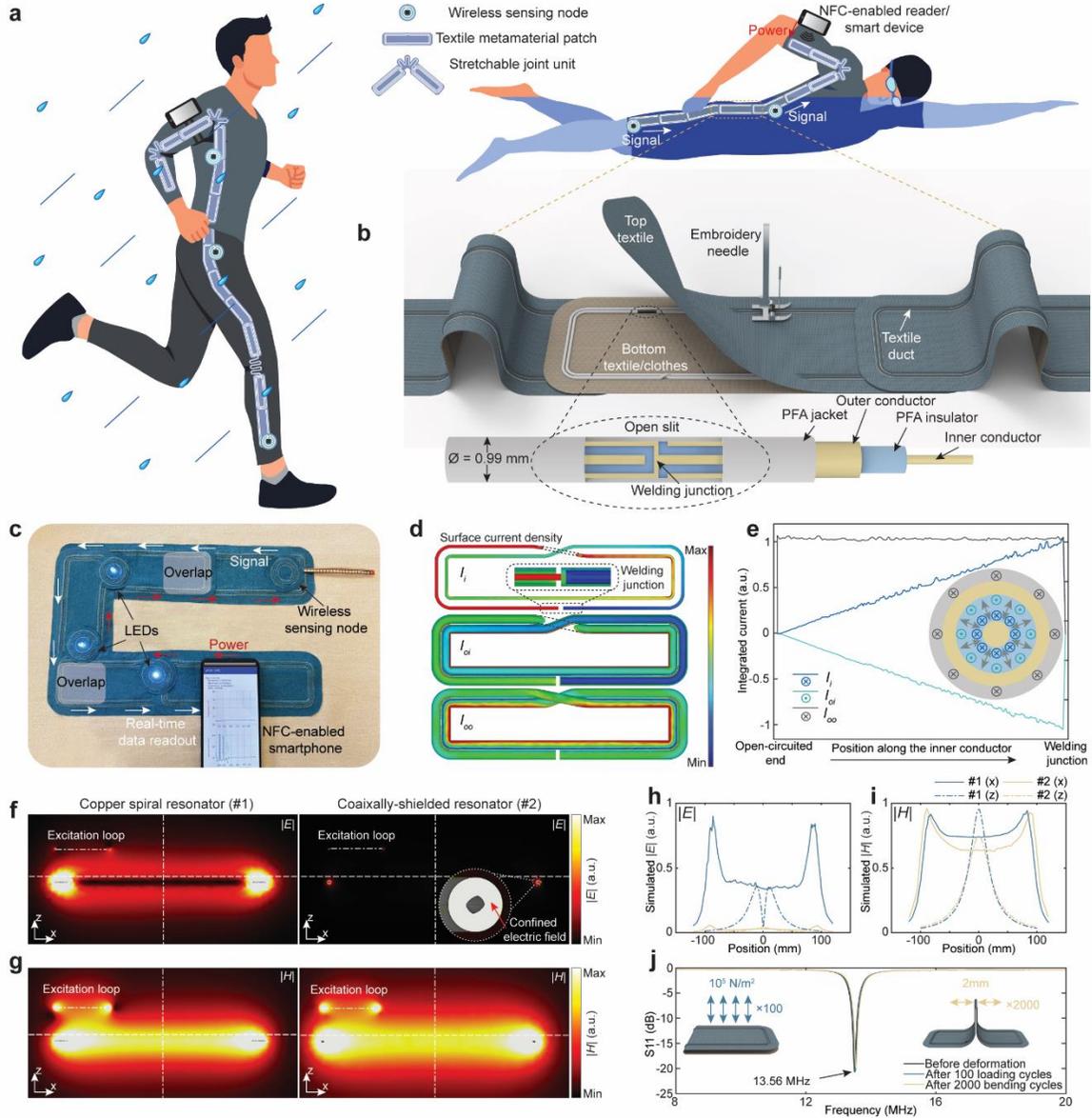

**Figure. 1 Coaxially-shielded textile metamaterial enabled robust near-field BAN. a** Concept of near-field BAN-integrated smart clothing, capable of interconnecting multiple sensing nodes while remaining robust against rainy environment and water. **b** Digital embroidery process and the sectional view of the open slit inside the coaxially-shielded metamaterial unit cell. **c** A 5-unit metamaterial array being used to power three LED nodes and a wirelessing sensing node simultaneously through an NFC-enabled smartphone. **d, e** Simulated surface current density on the associated conductor surface (dimension is not to scale) inside a metamaterial unit cell (**d**) and the integrated surface current profile along the metamaterial conductor (**e**). Inset: Cross-sectional view of the coaxial cable where the current is integrated along the azimuthal direction. Grey arrows denote the electric field confinement. **f, g** Simulated side view (longer side) of the electric field (**f**) and magnetic field (**g**) distribution in the vicinity of a copper spiral resonator and a coaxially-shielded resonator, sharing the same dimension and tuned to 13.56 MHz. **h, i** Electric field (**h**) and magnetic field (**i**) plotted along the white dashed lines in **f, g**. **j** The metamaterial's resonant frequency remains stable after mechanical deformations.

Typical functions of magnetic metamaterials, such as field enhancement and phase modulation, are primarily achieved through inter-unit cell coupling and the resulting collective resonant modes[33-36]. An inter-unit cell coupling can be created by introducing adequate overlap in the constructed metamaterial array, allowing it to be freely extended to form a transmission relay along which the NFC signal can be transmitted through inductive coupling in the form of a magneto-inductive wave[37,38]. While magneto-inductive wave can also propagate along three-dimensional coaxially-aligned inductively coupled loop arrays[39,40], the in-plane arrangement adopted herein provides higher mechanical flexibility and extendibility, better conforming to the irregular shape of human anatomy. When an NFC reader or smartphone is placed in proximity to one unit cell in the network, the reader's command can be broadcasted along the relay to power an energy harvesting node (e.g., light emitting diodes (LEDs)) or establish communication with a wireless sensing node within the network (Fig. 1c).

The resonance of the metamaterial is exclusively determined by its inherent geometric structure. Upon excitation, a resonating current ($I_i$) is generated on the surface of the inner conductor, traversing the entire inner conductor while linearly increasing until it reaches a maximum value at the welding junction. Simultaneously, a mirrored inductive current ($I_{oi}$) with the same magnitude but opposite direction emerges on the inner surface of the outer conductor due to the close distance between the two conductor layers (0.13mm). At the welding junction, the inner and outer conductors are not only directly welded but also connected by a parasitic loop through the outer surface of the outer conductor, along which a third current ($I_{oo}$) uniformly flows in the same direction as $I_i$ due to the skin depth effect. Figure 1d depicts the simulated surface current density distribution on their respective conductor surfaces. These currents are subsequently integrated along the azimuthal direction on the cable's cross-sectional plane (Fig. 1e, Supplementary Fig. 2a, d). Notably, $I_i$ and $I_{oi}$ have the same amplitude and opposite direction throughout the metamaterial unit cell, indicating that when viewed from a distance, they effectively cancel each other in terms of their inductive magnetic field, leaving $I_{oo}$ as the sole contributor to the propagating magneto-inductive wave in the network. In fact, the behavior of

$I_i$ and $I_{oi}$ is similar to the concept of a differential mode signal, carrying the transmission signal inside a coaxial cable when it functions straightforwardly as a transmission line[41]. The coaxial cable serves to shield the signal from outside perturbation or electromagnetic interference. The behavior of $I_{oo}$, on the other hand, resembles the common mode signal, usually induced by extraneous perturbations as noise that travels on the outer surface of the coaxial cable. In various RF scenarios, the common mode signal is undesired and often suppressed by a balun circuit or a differential amplifier. However, in the case of the metamaterial presented herein, the current $I_{oo}$ is intentionally induced and serves as the contributor for the magneto-inductive wave, carrying the NFC signal.

Importantly, the potential difference between the inner and outer conductor gives rise to an electric field confinement effect between the two conductor layers, resulting in a substantial attenuation of the electric field in the proximity of the metamaterial (Fig. 1f, h), where the presence of parasitic capacitance from human anatomy and environmental variations induces minimal and even negligible capacitive coupling, eventually contributing to the improved spectral stability of the metamaterial. The electric field confinement also indicates the formation of a significant structural self-capacitance between the two conductor layers, enabling a lumped element-free design while maintaining resonance at the NFC working frequency of 13.56 MHz, despite its relatively small dimension compared to the NFC wavelength (22 m). Additionally, this electric field confinement effect does not adversely impact the metamaterial's predominant magnetic resonance, resulting in a comparable inductive magnetic field to conventional copper spiral-based metamaterial (Fig. 1g, i). The latter lacks an effective electric field confining mechanism, making it sensitive and spectrally unstable to extraneous loading. We fabricate a textile metamaterial patch unit with the proposed technique and tune it to 13.56 MHz. The metamaterial unit cell also remains stable after mechanical deformations, including overall vertical pressure and cyclic bending, as shown in Fig. 1j.

**Electromagnetic characterization of the BAN**

As previously mentioned, introducing adequate overlap enables inter-unit cell coupling, facilitating the propagation of the magneto-inductive wave along the metamaterial array. The flexibility and mechanical stability of the metamaterial allow for its free extension and population along the irregular shape of various clothing. However, as more units are integrated, signal attenuation becomes more predominant at the array terminal. To maintain the robustness of the textile metamaterial-enabled BAN under various user-defined configurations, determining an optimal overlap, or optimal inter-unit cell coupling is crucial.

The inter-unit cell coupling is predominantly determined by the overlap of the magnetic flux while on resonance. As the geometry of each unit cell is identical, the coupling coefficient between two adjacent unit cells can be expressed as $k = L_M/L_S$, in which $L_M$ represents the mutual inductance between the unit cells, and $L_S$ represents the self-inductance of a single unit cell. Employing electromagnetic theory, we can define the effective mutual inductance and self-inductance of discrete unit cells as[42]:

$$L_{ij} = \frac{\mu_0}{4\pi |I_i I_j|} \iint dr_i r_j \frac{J(r_i) \cdot J(r_j)}{|r_i - r_j|} \tag{1}$$

where $r_i$ and $r_j$ are the integration elements along the current path of the metamaterial unit cell, representing the outer surface of the outer conductor since the contribution to inductance from $I_i$ and $I_{oi}$ will cancel each other. $J(r_i)$ and $J(r_j)$ are the spatial current densities at $r_i$ and $r_j$, respectively, maintaining a constant value according to Fig. 1e. $I_i$ and $I_j$ are the magnitude of the equivalent electric currents on the unit cell, equal to the magnitude of $I_{oo}$. Equation (1) denotes the self-inductance when $i$ and $j$ are equal and mutual inductance when $i$ and $j$ differ. Upon introducing the inter-unit cell area overlap $\alpha$, the coupling coefficient $k$ between two unit cells is numerically calculated and plotted as a function of $\alpha$ in Fig. 2a. When two identical unit cells are placed adjacent to each other, a resonant mode splitting can be expected, and the resulting two modes may be utilized to retrieve the coupling coefficient according to[43]:

$$k = \frac{f_o^2 - f_i^2}{f_o^2 + f_i^2} \tag{2}$$

in which $f_o(f_i)$ represents the resonant frequency of the mode with opposite (identical) current direction in the two unit cells after mode splitting. The coupling coefficient between two unit cells is subsequently measured and plotted in Fig. 2a, demonstrating a high degree of agreement with the theoretically derived value. It should be noted that the current direction in each unit cell does not affect the transmission of the magneto-inductive wave or the reader-tag communication.

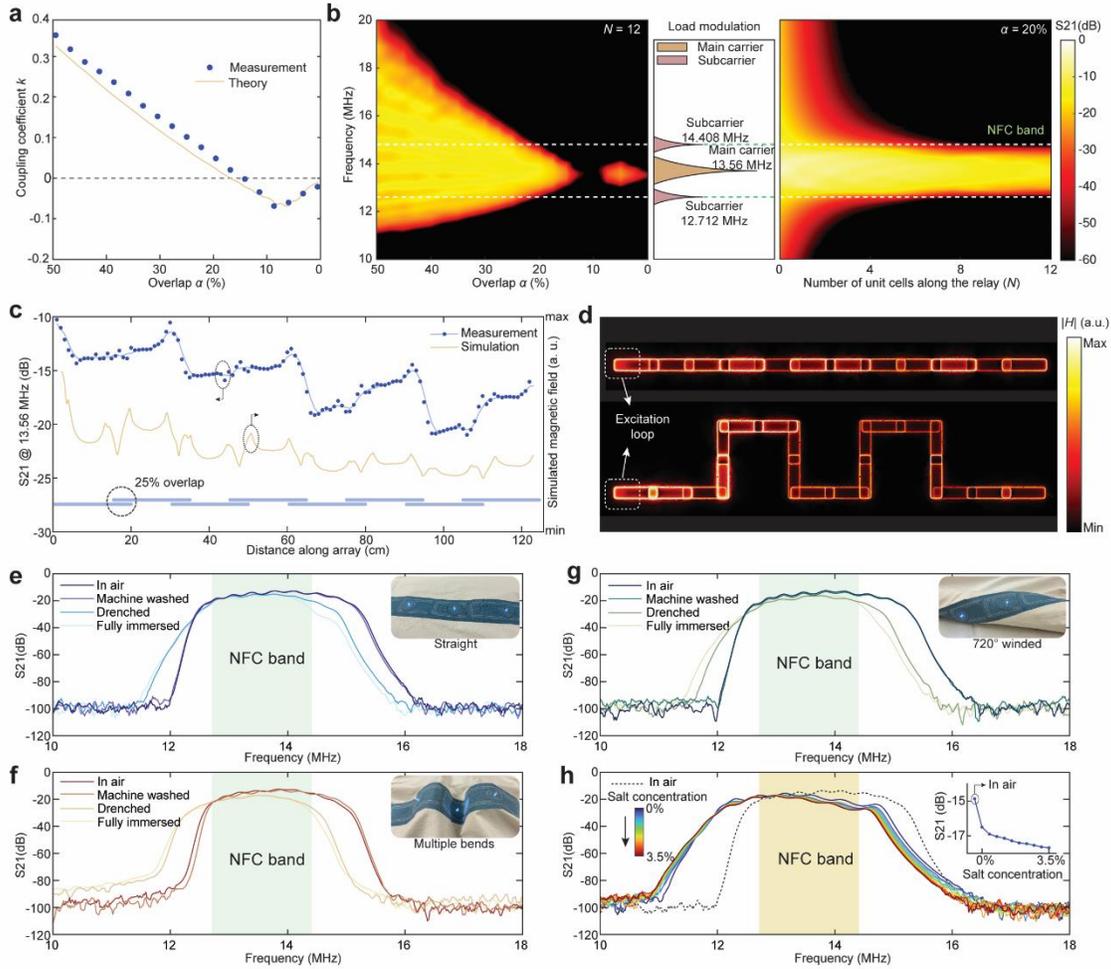

**Figure. 2 Electromagnetic characterizations of the BAN. a** Measured inter-unit cell coupling coefficient ($k$) and the theoretically derived value as a function of inter-unit cell overlap ($\alpha$). **b** Measured transmission coefficient of BANs with different $\alpha$ and unit numbers ($N$). The NFC load modulation requires an NFC passband that can be used to leverage an optimal $\alpha$ value. **c** Measured transmission coefficient along a 12-unit inline array and the simulated magnetic field strength. **d** Simulated magnetic field distribution at 13.56 MHz along two arrays, showing robust signal transmission. **e-g** The 12-unit array demonstrates spectral stability after machine washing, being drenched, or entirely immersed in DI water when configured straight (**e**), with multiple bending (**f**), and 720° wound (**g**). Insets: Pictures of the textile metamaterial array being deformed while powering three battery-free LED nodes. **h** The BAN is insensitive to loss induced by conductive matter, covering the NFC passband when submerged in saline solutions. Inset: S21 measured at 13.56 MHz with increasing salt concentration.

The performance of signal transmission in the textile metamaterial-enabled BAN is then characterized using an inline array composed of $N$ = 12 unit cells. The transmission coefficient is measured using a transmitter and a receiver, both coupled to a vector network analyzer (VNA) and placed at each terminal of the array, as a function of $\alpha$ (Fig. 2b). For a larger $\alpha$ (larger $k$), the original resonant mode at 13.56 MHz splits into $N$ modes that form a collective passband centered at the original mode (Supplementary Fig. 3), as indicate by equation (2). A larger $k$ allows for the magneto-inductive wave to pass along the array with mitigated loss, but it also gives rise to a more predominant mode splitting, resulting in a wider band (Supplementary Fig. 4). This excessive bandwidth causes notable fluctuations in the magnetic field distribution along the array, requiring more units for the same transmission distance and eventually resulting in higher attenuation at the end of the array. For smaller $\alpha$ (smaller $k$), or the most extreme case, critical overlap[44], which results in zero value of $k$ (maps to $\alpha \approx$ 10% in Fig. 2a), the unit cells are loosely coupled and introduce substantial attenuation to the NFC signal (Supplementary Fig. 4), which is also not ideal. The NFC readers typically broadcast their read command in the sidebands of the main carrier located at 13.56 MHz, but the load modulation, which varies the impedance of the NFC tag antenna, is carried in the sidebands of two auxiliary subcarriers located 848 kHz from the main carrier[45]. This requires the metamaterial to cover a bandwidth of at least 1.7 MHz centered at 13.56 MHz, which maps to $\alpha$ = 25% and $k$ = 0.1 according to Fig. 2a. The manipulatable bandwidth of the metamaterial by varying the overlap can be leveraged to further increase robustness against spectral shift at the expense of sensitivity.

The transmission coefficient of the metamaterial was then measured by integrating various numbers of unit cells in an inline array ($N$ = 1 to 12), while maintaining $\alpha$ = 25%. The array can cover a reasonably long distance up to 185cm, sufficient to populate common clothing around the human anatomy. The transmission coefficient along the array, characterizing the magnetic field strength, was measured by fixing the VNA's transmitter loop at one end and moving the receiver loop along the metamaterial (Fig. 2c). Due to the formation of phase transition regions, especially at positions where adjacent unit cells overlap, the transmission coefficient profile shows fluctuations. The magnetic field was also derived from simulation and plotted in Fig. 2c, demonstrating good agreement with both the measured result and the pattern of the metamaterial alignment. The measured transmission profile shows less fluctuation than the simulation results, as the field is averaged within the VNA receiver loop.

Figure 2d displays the simulated magnetic field distribution along a 12-unit inline array and a 20-unit convoluted array, demonstrating the robustness of signal transmission in the metamaterial and the capability of the metamaterial to be extended along various directions. According to Fig. 2b, the inter-unit cell coupling gradually approach 0 as the adjacent unit cells are separated. In an arrangement where $\alpha$ = 25% is adopted, the resonance of the metamaterial is predominantly determined by the nearest neighbor coupling, and the next nearest neighbor coupling may be neglected. Therefore, as long as the $\alpha$ value is maintained, the inductively coupled metamaterial may rotate freely for better conformance to clothing and be extended to the desired sensing node locations (Supplementary Fig. 5).

The performance of the metamaterial is further examined against more severe conditions involving mechanical deformation, such as random bending and winding, which represent common deformations of clothing in daily use cases. Even under these mechanical deformations, the metamaterial demonstrates significant spectral stability against various loading conditions (Fig. 2e-g). The strong electric field confinement from the coaxially-shielded structure contributes to the metamaterial's robustness, showing a frequency deviation of less than 4% for the transmission band when drenched or completely immersed in deionized (DI) water. The transmission profiles indicate full coverage of the NFC passband even when the metamaterial is completely surrounded by water, which has a relative permittivity as high as 80 (as opposed to typical loosely one-side loading of the human anatomy, with a permittivity averaging 40 - 60). Furthermore, the metamaterial, shielded with a PFA material layer and sealed with a waterproof heat shrink at the welding junction, maintains robustness against short-circuits. The metamaterial remains functional after machine washing, and proves insensitive to conductive matter. The transmission coefficient of the 12-unit inline array is measured when fully immersed in a saline solution with increasing salt concentration up to 3.5%, equivalent to the level of seawater and much higher than human sweat and natural raindrops. While the salty water may introduce a transmission loss with its increased electric conductivity, only a 2.8 dB attenuation is observed at 13.56 MHz for the peak transmission coefficient of the inline array, and the NFC band remains covered by the metamaterial's bandwidth (Fig. 2h). This result highlights the potential application of the coaxially-shielded metamaterial BAN not only in wet environments, including rain, perspiration, and showering but also proves valuable in a body-tracking system for aquatic sports, overcoming challenges posed by the high conductivity and transmission loss in seawater.

**Characterization of the stretchable joint unit**

The metamaterial, when integrated into textile layers, exhibits outstanding flexibility and can be seamlessly integrated onto pre-existing clothing, withstanding mechanical deformation resulting from natural human motion. However, the micro coaxial cable employed to craft the metamaterial inherently has low pulling tension and elastic strain, resulting in low stretchability. This limitation may hinder its applicability to cover joint positions in clothing, particularly concerning the shoulder joint with three degrees of freedom. The shoulder joint experiences wide and random flexion during human motion and exercise, leading to potential concerns about the stability of magneto-inductive wave transmission along the metamaterial, especially with random overlap or creasing of clothing at such joints. This could potentially interrupt continuous signal monitoring.

Consequently, a stretchable joint unit is designed, featuring a stretchable serpentine link connecting two separate 3-turn coaxially-shielded resonators that adopt an alternative internal topology, analogous to the regular metamaterial unit cell (Fig. 3a, supplementary Fig. 1b). This stretchable join unit can effectively replace regular unit cells in the metamaterial to cover joint positions while providing stable signal transmission regardless of the degree of joint flexion (Supplementary Fig. 6a, b). Each resonator in the joint unit features an outer cut where the outer conductor shield is intersected and left hanging, and an inner cut on the opposing side where the inner conductor is intersected. At the inner cut, one end of the inner conductor is left open-circuited for resonating current excitation, while the other end is electrically connected to an identical resonator through the serpentine link. Figure 3b shows the equivalent circuit diagram for the stretchable joint unit, with two identical resonators interconnected through the serpentine link. The frequency response of the fabricated joint unit is measured at different rotating angles, as depicted in Fig. 3c. The joint unit features two resonant modes, and neither mode is affected by the variation of the rotating angle. The working mode (left peak tuned to 13.56 MHz) demonstrates a negligible 0.34% frequency shift (0.046 MHz) during the rotation.

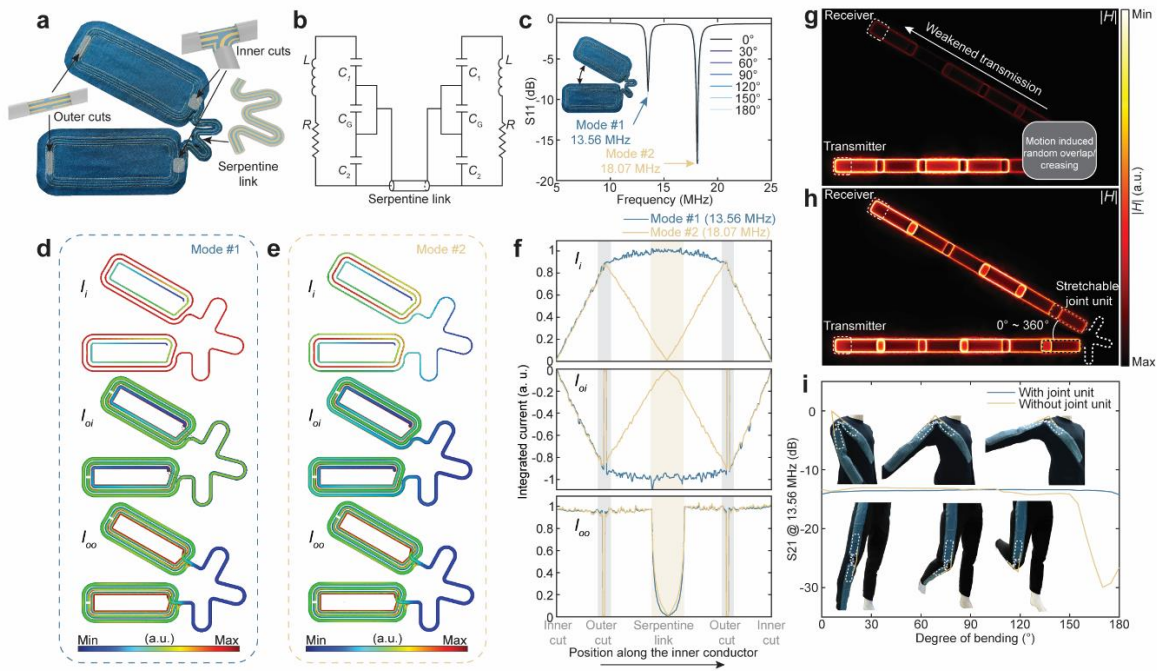

**Figure. 3 Increased stretchability with the joint unit. a** Picture of the stretchable joint unit that adopts an alternative internal topology. **b** Equivalent circuit diagram of the joint unit. $L$ and $R$ represent the inductance and series resistance of the inner conductor on each side of the serpentine link, $C_G$ represents the capacitance formed at the gap of the outer cuts, $C_1$ and $C_2$ represent the large structural capacitance formed between the outer conductor and the inner conductor at each side of the outer cuts. **c** Measured reflection coefficient of the joint unit under different rotating angle. **d-f** Simulated surface current density on the associated conductor surface (dimension is not to scale) for the working mode (**d**) and the higher second mode (**e**), and the integrated surface current profile along the conductor of the joint unit (**f**). **g, h** Simulated magnetic field distribution at 13.56 MHz showing robust signal transmission passing a bent joint unit (**g**), in contrast to weakened transmission when bending an array without a joint unit (**h**). **i** Measured transmission coefficient at 13.56 MHz of two 12-unit metamaterial arrays (with and without replacing the sixth and seventh units with a joint unit) as a function of the degree of bending. Insets: Pictures showing improved stretchability at shoulder and knee joints.

Upon excitation, the behavior of currents in the joint unit follows a similar pattern as the regular metamaterial unit. Specifically, $I_i$ and the induced $I_{oi}$ have the same amplitude but opposite direction, while $I_{oo}$ uniformly traverses the outer surface of the outer conductor, except for the segment of the serpentine link, as depicted by the simulation result of the surface current density in Fig. 3d, e. The working mode is distinguished from the second mode at 18.07 MHz by the fact that for the working mode, $I_i$ and $I_{oi}$ reach the maximum value at the outer cut and maintain this value through the serpentine link (Fig. 3f, Supplementary Fig. 2b, e, f), indicating that the two resonators are electrically connected to allow transmission of the NFC signal. In contrast, for the second mode at 18.07 MHz, all three currents drop to zero at the link, indicating no signal transmission between the two resonators at this frequency. In fact, the second mode represents the resonant mode when half the joint unit is operating alone without the link and the other half.

The simulated magnetic field distribution after bending a 12-unit inline metamaterial array with and without the stretchable joint unit shows that without the joint unit, the sixth and seventh units in the array may be subject to random additional overlap or creasing of the clothing induced by joint flexion, thereby weakening the transmitted signal (Fig. 3g). Conversely, when the sixth and seventh units are replaced with a joint unit, the transmission remains robust along the entire array (Fig. 3h). The measured transmission coefficient at 13.56 MHz of two 12-unit metamaterial arrays as a function of the degree of bending in the middle of the array (between the 6th and 7th unit) shows that the array equipped with the joint unit shows steady transmission regardless of the bending degree, while the one without the joint unit exhibits fluctuation in the transmission profile when the bending is beyond 100 degrees, and the transmission coefficient drops significantly when the bending is beyond 150 degree, as a result of introducing excessive overlap at the bending location (Fig. 3i, supplementary Fig. 6c, d).

**Construction of NFC tags using coaxially-shielded antenna**

For the coaxially-shielded resonator configuration where both an inner cut and an outer cut are present (Fig. 3a), the open-circuited inner conductor at the inner cut may be exposed and hardwired to a direct RF feed, an NFC transponder, or a load (e.g., LEDs), functioning as antenna with improved mechanical flexibility and spectral stability (Fig. 4a, Supplementary Fig. 2c, g). Employing the same embroidery technique used for the construction of the textile metamaterial patch unit, we create textile battery-free NFC tags, incorporating coaxially-shielded antennas and off-the-shelf NFC transponders integrated with a 14-bit sigma-delta analog-to-digital converter (ADC). The ADC interfaces with a thermistor and a strain sensor (or a variety of analog sensors), as shown in Fig. 4a and Supplementary Fig. 7. NFC tags fabricated in this manner are small, lightweight, flexible, and can be directly detected by smartphones or commercial NFC readers without addition components.

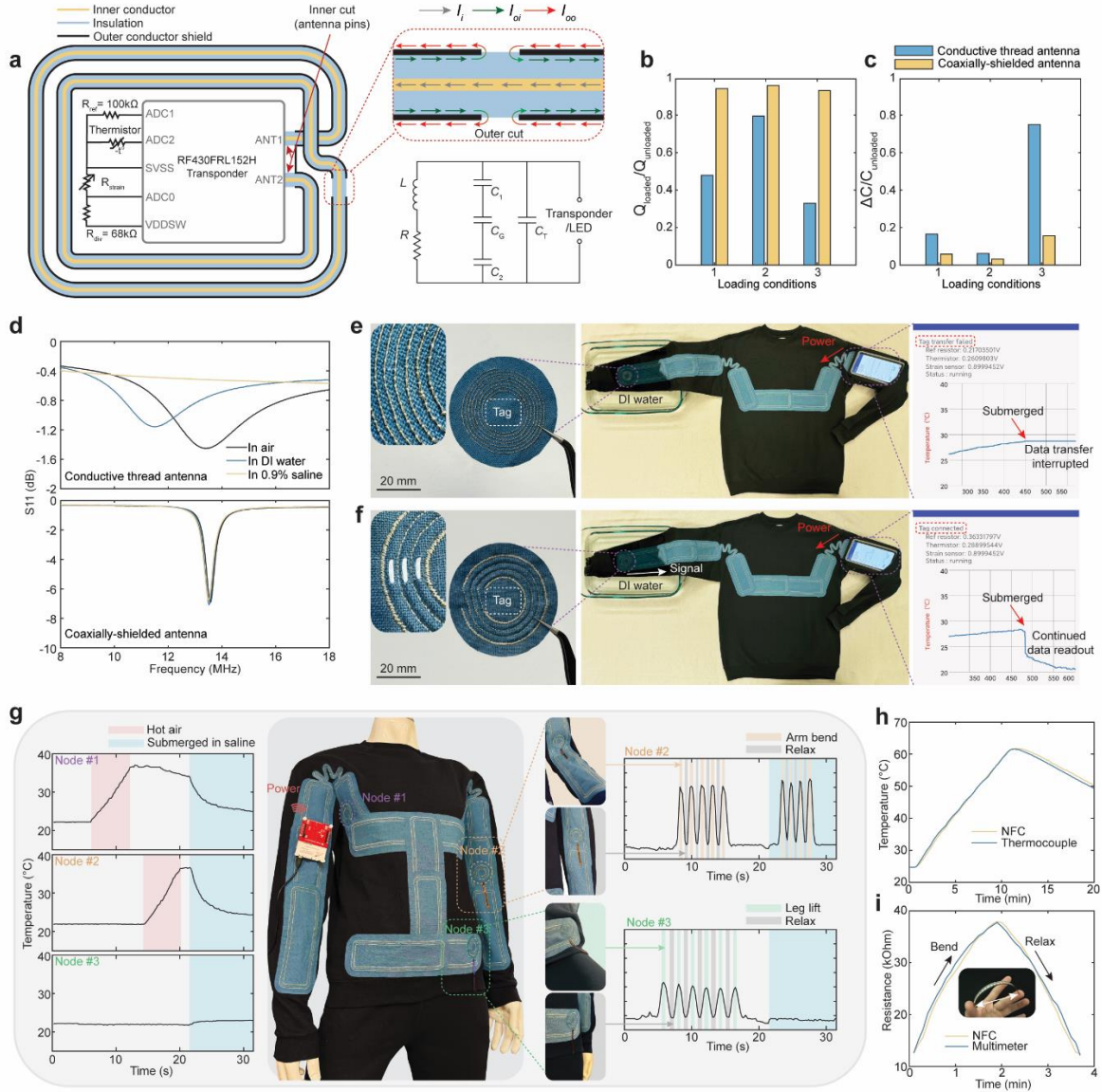

**Figure. 4 Construction of the BAN using NFC transponders with coaxially-shielded antennas. a** Simplified equivalent circuit diagram of an NFC tag integrated with a coaxially-shielded antenna, the current distribution at the antenna's outer cut and equivalent circuit diagram of the antenna. *L* and *R* represent the inductance and series resistance of the inner conductor, $C_G$ represents the capacitance formed at the gap of the outer cut, $C_1$ and $C_2$ represent the structural capacitance formed between the outer conductor and the inner conductor at each side of the outer cut. $C_T$ is the tunning capacitor. **b, c** Measured quality factor preservation ($Q_{Loaded}/Q_{Unloaded}$) (**b**) and self-capacitance variations $\Delta C/C_{Unloaded}$ (**c**) of the two antennas when affixed to skin (1), wetted by sweat (2) and submerged in water (3). **d** Measured reflection coefficient of the two antennas. **e, f** NFC tag with coaxially-shielded antenna promotes continuous data readout during submersion in water (**e**) while data transfer through the conductive thread antenna is interrupted (**f**). **g** Proof of concept demonstration of multi-node signal monitoring through textile metamaterial BAN-integrated clothing. The sensor data is simultaneously recorded by three battery-free wireless sensing nodes. **h, i** Calibrations of the thermistor using a wired thermocouple (**h**) and the strain sensor using a multimeter (**i**).

Conductive thread is a commonly-adopted viable option for constructing antennas for wireless communication through digital embroidery. We design both a coaxially-shielded antenna and an alternative conductive thread antenna, both optimized to share the same coverage and filling ratio while operating at 13.56 MHz for NFC communication. The coaxially-shielded antenna exhibits the highest quality factor of $Q_{unloaded} = 33$ at 13.56 MHz, which is 6 times higher than that of the conductive thread antenna with $Q_{unloaded} = 5.3$. This discrepancy is primarily attributed to the low resistance (0.9 Ωm$^{-1}$) of the copper-based inner conductor core in the coaxial cable compared to the higher resistance of the lossy commercial conductive thread (50 Ωm$^{-1}$). The coaxially-shielded antenna also exhibits robust quality factor preservation ($Q_{loaded}/Q_{unloaded}$) when compared to the conductive thread antenna (Fig. 4b) and a commercially available flexible NFC antenna with similar dimensions (Supplementary Fig. 8), retaining more than 90% of the quality factor when affixed to skin, wetted by sweat, or submerged in water (Fig. 4b). In contrast, the conductive thread antenna, despite being sealed in an insulating filming dressing (see Methods section), maintains only 48%, 79% and 33% of its maximum quality factor under these conditions. The conductive thread antenna also exhibits significant self-capacitance variations under these loadings (Fig. 4c), measured at the antenna pins using an impedance analyzer, indicating its susceptibility to dynamic environmental variations. This susceptibility is further evidenced by a 2 MHz shift in its working frequency when in contact with DI water and a short-circuit occurrence when exposed to a saline solution (Fig. 4d). The coaxially-shielded antenna, in sharp contrast, shows only a negligible 0.59% frequency shift (0.08 MHz) when immersed in the saline solution, demonstrating robustness against both dielectric loss and conductive loss. While commercially available conductive thread with encapsulation may mitigate the risk of short circuits and reduce the conductive loss[24], its high conductor resistance and the spectral instability induced by dielectric loss persist, and the associated increase in stiffness and dimension may prove incompatible with the adopted embroidery machine. Additionally, recent developments in NFC antenna design leverage advanced conductive material properties and fabrication techniques, proposing various NFC antenna designs including stacked multilayer graphene-based traces[46], digitally embroidered conductive yarn[47], and 3D-printed liquid metal microchannels[48]. However, beyond potential fracture, leakage, or biocompatibility issues, a shared limitation among these innovations is their limited spectral stability—a crucial characteristic of antennas that can significantly affect signal transmission.

We then construct battery-free NFC tags adopting these two antennas, and their data transfer capabilities is investigated through a shirt integrated with a metamaterial BAN. The tag is power by an NFC-enabled smartphone with a developed custom Android application that retrieves real-time sensor data from the tag. While both tags can be detected through the metamaterial network, data transfer through the conductive thread antenna is interrupted when the tag is immersed in DI water (Fig. 4e). The coaxially-shielded antenna, in contrast, facilitates uninterrupted data readout during the submersion process (Fig. 4f).

**Proof-of-concept multi-node signal monitoring**

The underwater signal transmission capability of the coaxially-shielded antenna complements the resilience of the textile metamaterial relay in wet environments. The combination of the textile metamaterial patch unit, stretchable joint units and textile NFC tags enables the creation of highly customizable near-field-enabled clothing tailored to user preferences. We design a complete textile metamaterial-enabled BAN on a long-sleeve sweatshirt, as illustrated in Fig. 4g. By maintaining the pre-determined 25% overlap between adjacent metamaterial unit cells, the design of specific metamaterial-enabled clothing is highly flexible, and the network can be freely extended, routed, and branched into separated pathways across the clothing to reach the desired sensing node locations (Supplementary Fig. 9). Two stretchable joint units are strategically placed at the shoulder joints to enhance stretchability.

To demonstrate the potential applications of the metamaterial BAN in physiological signal monitoring, we implement a multi-sensor readout as a proof of concept. Three battery-free sensing nodes are placed under the armpit, on the elbow, and at the waist region, each equipped with a thermistor to monitor peripheral temperature variations in three distinct temperature zones, including the axillary temperature —a crucial indicator of various health conditions. The sensing nodes on the elbow and waist also include a strain sensor for potential gait monitoring during exercise such as running or swimming. Powering of the tags is achieved by placing a commercial NFC reader, powered by a laptop through a USB cable, on the arm— a configuration analogous to how smartphones are typically affixed with an armband during exercises.

During the sensor data recording period, the environmental temperature is maintained at 22 °C, the sensing nodes 1 and 2 are heated by a heat gun for a short period of time to simulate the elevation in human body temperature during exercise, followed by a natural cooling phase. The network's capability to maintain connectivity between the reader and tags under a lossy loading condition is demonstrated by subsequently submerging the entire network into a saline solution (0.9% salt concentration, 24°C). This induces a rapid cooling of the heated thermistors, with the thermistors accurately capturing the sudden temperature drop. Additionally, bending is also performed on sensing nodes 2 and 3 to simulate arm and leg movement during a short-period exercise. Despite the BAN and sensing nodes undergo a sudden transformation in the surrounding environment, characterized by the presence of conductive matter and a significant increase in the relative permittivity of the surrounding media, the power supply and reader commands consistently reach each sensing node without interruption. We then calibrate the adopted thermistor using a data logger wired to a thermal couple (Fig. 4h), and the strain sensor is calibrated using a multimeter (Fig. 4i), the data measured with the NFC tag agrees well with the calibration measurements.

This continuous sensing capability holds potential to track vital signals during athletic activities involving water or substantial perspiration, environments where dynamic conditions prevail. Monitoring key indicators such as heart rate and respiratory rate assists participants in performance evaluation and regulation of activity intensity, thereby mitigating the risk of overexertion and enabling the early detection of distress.

**Discussion**

We present an innovative metamaterial design tailored for the creation of a scalable BAN with enhanced near-field capabilities. This network holds promise for applications in health monitoring and body-tracking across various activities, including aquatic sports, where the network may encounter a conductive and lossy environment. The metamaterial, crafted through computer-aided digital embroidery into a flexible and durable textile patch, can be applied onto pre-existing clothing in a highly customizable configuration. Leveraging the layered internal structure of the metamaterial, the electric field is effectively confined within the coaxial cable, mitigating capacitive detuning induced by environmental perturbations. Despite its small size and absence of lumped elements, the metamaterial exhibits strong magnetic resonance at the NFC working frequency of 13.56 MHz, enabling robust magnetic coupling between adjacent unit cells. This inductive coupling allows an NFC polling device, such as a commercial reader or smartphone, to both power and establish communication with battery-free wireless sensing nodes within the network, continuously retrieving sensor data from them. A proof-of-concept demonstration of a simultaneous multi-sensor readout underscores the potential of the proposed metamaterial, along with the developed textile NFC tags, for achieving robust continuous signal monitoring at a system level, especially in challenging environmental conditions.

The construction method for the BAN involves embedding the metamaterial as a conductive interlayer into clothing. In comparison to previously suggested approaches[24,26], this method eliminates the direct use of conductive wires to interconnect multiple sensor locations. The metamaterial patch unit can be constructed individually and then bonded onto clothing using a heat-fusible tape, avoiding an irreversible embroidery process while providing users with maximum flexibility in sensor placement, as shown in the alternative metamaterial patterns involving various sensing functionalities (Supplementary Fig. 10). Our metamaterial design intentionally introduces a substantial structural capacitance to bind the electric field within the layered internal structure, imparted by industrially certified coaxial cable. This simple, straightforward design minimizes undesired interference with the surrounding environment with minimal effort, while eliminating lumped elements and fragile internal structures proposed elsewhere[25,49].

The selection of the micro coaxial cable (diameter 0.99mm) balances ease of fabrication and flexibility, resulting in a lightweight, unperceivable, and comfortable BAN for the user. Further improvements in invisibility may be achieved with smaller-diameter industrial micro coaxial cable solutions (as small as 160μm), potentially enabling direct embroidery onto clothing (Supplementary Fig. 11) and serving as a textile transmitter for wirelessly powering biomedical implants.

Future work may involve integrating advanced sensing modalities, extending the application to long-term physiological signal or gait monitoring at the clinical trial level, paving the way for a BAN-enabled personalized and proactive healthcare system, as well as a body-tracking system for various sports. Additionally, adapting the BAN to the unique requirements of highly stretchable sports apparel, such as swimwear, track suits, or gymnastic uniforms, may involve creating a coaxially-shielded metamaterial incorporating layered stretchable insulating elastomer and conductive fiber. Addressing signal transmission among separate near-field-enabled clothing is also a consideration for future research. While the current BAN allows signal transmission through a discontinuous gap between adjacent clothing, this connection hinges on the relative positions of the two BANs and their loosely inductive coupling (Supplementary Fig. 12), making it susceptible to human motion. A potential solution involves combining NFC with other radiative wireless communication protocols, equipping each clothing item with a wireless communication module to establish a more stable wireless connection across discontinuities.

## Methods

**Metamaterial construction.** The metamaterial patch unit was constructed from off-the-shelf micro coaxial cable (9434, Alphawire). A computer-aided digital embroidery pattern, designed to conform to the metamaterial contours, was created using commercial software (Hatch Embroidery 3, Wilcom). The embroidery process was carried out by a digital embroidery machine (PE535, Brother), securely bonding two layers of thin, lightweight textile (50% linen, 50% cotton), resulting in the formation of a textile duct where the metamaterial was then assembled. The stretchable join unit was constructed in a similar approach. The transfer of the metamaterial onto clothing or substrate textile were achieved through a heat-fusible hemming tape, which securely bond the metamaterial to its substrate after heat pressing.

**NFC tag construction.** The miniaturized NFC tag circuit was patterned on a FR-4 substrate using a printed circuit board (PCB) prototype machine (ProtoMat S64, LPKF), with the board measuring 12 mm × 16 mm. The NFC tag incorporated a commercial NFC transponder (RF430FRL152H, Texas Instrument (TI)) implementing the ISO 15693 protocol. The transponder featured a 14-bit ADC that interface with a commercial negative temperature coefficient thermistor (B57471V2104J062, TDK Electronics) with a beta value of 4480 $K$ and a strain sensor (long flex sensor, Adafruit). The transponder was wired to a 3-turn coaxially-shielded antenna (outer diameter: 40 mm, inner diameter: 20mm) optimized to better couple to the BAN, the programmable digital embroidery pattern can be leveraged for alternative antenna patterns for different purposes (Supplementary Fig. 13). The antenna was constructed similarly to the metamaterial and integrated into two layers of textile. The conductive thread antenna was formed by embroidering a commercial 2-ply conductive thread (Adafruit) composed of stainless-steel fibers. The energy harvesting node, which serves as an indicator of successful wireless communication, was constructed by directly wiring the coaxially-shielded antenna to an LED. The weight of a constructed sensing node is 3.8 g.

Measurement of the quality factor and self-capacitance was achieved by an impedance analyzer (IM7581, Hioki). When measuring these parameters, the entire conductive thread antenna was sealed in a polyurethane film dressing to prevent short circuits. For a fair comparison of how dielectric loss and conductive loss impact the quality factor of both antennas, only the exposed pins of the coaxially-shielded antenna were sealed. When measuring their resonant frequency, only the exposed pins of both antennas were sealed. Affixing the antenna to human skin was simulated by using a 3D-printed phantom with 1% agarose gel in 1 X phosphate buffered saline.

Calibration of the thermistor was achieved by placing a thermal couple (wired to a temperature data logger) in contact with the thermistor (mounted on the NFC tag). Both were affixed to a hot plate, heated for 10 minutes, followed by natural cooling. The calibration of the strain sensor was achieved by mounting one end of the strain sensor on a fixed stage, while the other end was attached to a motorized stage that was programmed to introduce bending of the sensor. The resistance values were measured twice in a wireless approach (through NFC) and a wired approach (through a multimeter) separately.

**Numerical simulation.** All simulations were carried out using the frequency domain solver in CST Microwave Suite 2021. The transmitter and receiver loops were driven by a discrete port. Simulation models shown in Fig. 1d and Fig 3d, e are not to scale, a thicker cable diameter was adopted for better visualization of the surface current distribution. Simulation models for deriving the current distribution in Fig. 1e and Fig 3f are to scale.

**Bench measurement.** All measurements of the S-parameters were performed using 5 cm × 6 cm loop antennas coupled to a VNA (P5020B, Keysight Inc.). The loop antennas were placed directly on the metamaterial unit cell or the array, with no vertical separation distance. To measure the transmission coefficient along the metamaterial array, the transmitter loop was fixed at one end, while the receiver loop was mounted on a motorized linear stage programmed to move along the array.

For bending cycles carried out on a single metamaterial patch unit (Fig. 1j), a linear stage was employed. The stage was programmed to repeatedly compress the bent portion of the metamaterial to a minimal separation distance of 2 mm. Vertical loading of the unit was achieved by applying weight on top of the unit in a folded configuration. The frequency response of the unit was recorded both before and after the mechanical deformation.

**Continuous data readout.** Sensor data readout by the NFC-enabled smartphone was accomplished by a developed custom Android application, which can establish communication with one NFC tag and retrieve three 16-bit ADC values in real-time. The data was then displayed in the form of input voltages at the associated pins, subsequently converted to resistance of the temperature and strain sensors and plotted as a function of time.

Multi-sensor readout was achieved using a commercial NFC reader (DLP-7970ABP, TI) mounted on an MSP430G2553 micro controller (TI). The reader received power from a laptop through a USB cable. Communication between the reader and the tags implemented the ISO 15693 protocol. The three tags were strategically placed under the armpit, on the elbow, and on the waist, with wireless communication distances of 0.4 m, 1 m and 0.9 m along the metamaterial relay separately. The sampling rate of each sensor is 5 Hz during this process.

For all experiments involving submerging the NFC tag in DI water or saline solutions, the circuit and antenna pins of the tag were sealed by a polyurethane film dressing to prevent short circuits and protect the circuit components. However, the antenna remained entirely exposed to water without any sealing.

**Data availability**

The data that support the findings of this study are available from the corresponding author upon reasonable request.

**Acknowledgements**

This research was supported by the Rajen Kilachand Fund for Integrated Life Science and Engineering. The authors thank Boston University Photonics Center for technical support.

**Author contributions**

X. Zhu and K. W. contributed equally to this work. X. Zhu, and X. Zhang conceived the study. X. Zhu, K. Wu and X. Zhang designed and constructed the metamaterial. X. Zhu, K. Wu, X. Xie, and X. Zhang designed and conducted the bench measurements. X. Zhu, K. Wu, and X. Zhang conducted numerical modeling and theoretical analysis. All authors participated in discussing the results. X. Zhu, K. Wu, S. W. Anderson, and X. Zhang wrote the manuscript.

**Conflict of interest**

The authors declare no conflict of interest.

**Additional information**

Additional information is available for this work.